\documentclass[twocolumn,showpacs,preprintnumbers,amsmath,amssymb,pra]{revtex4}
\usepackage{graphicx}

\begin{document}
\thispagestyle{empty}
\title{Recent breakthrough and outlook in constraining the
non-Newtonian gravity and axion-like particles
 from Casimir physics}

\author{
G.~L.~Klimchitskaya}
%
%
\affiliation{
Central Astronomical Observatory
at Pulkovo of the Russian Academy of Sciences,
Saint Petersburg, 196140, Russia}
\affiliation{
Institute of Physics, Nanotechnology and
Telecommunications, Peter the Great Saint Petersburg
Polytechnic University, Saint Petersburg, 195251, Russia
}
%

%

\begin{abstract}
The strongest constraints on the Yukawa-type corrections to
Newton's gravitational law and on the coupling constants
of axion-like particles to nucleons, following from recently
performed  experiments of the Casimir physics, are presented.
Specifically, the constraints obtained from measurements of
the lateral and normal Casimir forces between sinusoidally
corrugated surfaces, and from the isoelectronic experiment
are considered, and the ranges of their greatest strength
are refined. Minor modifications in the experimental setups
are proposed which allow strengthening the resultant
constraints up to an order of magnitude. The comparison
with some weaker constraints derived in the Casimir regime
is also made.
\end{abstract}
\maketitle
\section{Introduction}

It is well known that many extensions of the Standard Model predict the
existence of light scalar particles \cite{1,2}. An exchange of one such
particle between two atoms results in the Yukawa-type correction to
Newton's gravitational potential \cite{3}. The same correction arises
in the multi-dimensional unification theories with a low-energy
compactification scale \cite{4,5}. It should be stressed that within the
micrometer and submicrometer interaction ranges such kind corrections
are consistent with all available experimental data even if they exceed
the Newtonian gravity by the orders of magnitude \cite{3}. However,
the gravitational experiments of Cavendish type \cite{6,7}, measurements
of the Casimir force \cite{8,9}, and experiments on neutron scattering
\cite{10} allow one to constrain parameters of the Yukawa-type interaction
in micrometer and submicrometer ranges.

An important object of the Standard Model and its generalizations is
the pseudoscalar particle axion introduced in \cite{11,12} for exclusion
of large electric dipole moment of a neutron and strong CP violation in
QCD. More recently, axions and various axion-like particles have been
actively discussed as the most probable constituents of dark matter
\cite{1,13}. Although an exchange of one axion between two nucleons
results in the spin-dependent potential, which averages to zero after
a summation over the volumes of two macroscopic bodies, an exchange of
two axion-like particles interacting with nucleons via the pseudoscalar
Lagrangian leads to the spin-independent force \cite{14}. Using this fact,
the coupling constants of axion-like particles to nucleons have been
constrained from the results of Cavendish-type experiments \cite{15,16}
and measurements of the Casimir-Polder and Casimir forces \cite{17,18,19,20}.

Recently, a considerable strengthening of constraints on the strength of
Yukawa interaction in the wide range from 40\,nm to $8\,\mu$m was achieved
in the so-called isoelectronic (Casimir-less) experiment, where the
contribution of the Casimir force was nullified \cite{21}. The obtained
constraints are up to a factor of 1000 stronger than the previously known
ones. The experimental data of \cite{21} were also used to strengthen the
constraints on the axion-to-nucleon coupling constants. The stronger up
to a factor of 60 constraints have been obtained over the wide range of
axion masses from 1.7\,meV to 0.9\,eV \cite{22}. This corresponds to the
wavelength of axion-like particles from $1.2\times 10^{-4}\,$m to
$2.2\times 10^{-7}\,$m, respectively.

This paper summarizes the strongest constraints on the Yukawa-type
corrections to Newtonian gravity and on the coupling constants of axion-like
particles to nucleons obtained so far from the Casimir physics.
It is demonstrated that by minor modifications of the already performed
experiments with retained sensitivity and other basic characteristics it is
possible to find even stronger constraints. Specifically, the experiment
\cite{23,24} is considered on measuring the lateral Casimir force between two
aligned sinusoidally corrugated Au-coated surfaces of a sphere and a plate.
It is shown that with appropriately increased corrugation amplitudes and
decreased periods the constraints of \cite{8} on the parameters of Yukawa-type
correction to Newtonian gravity can be strengthened by up to a factor of 10.
The same modification made in the experiment \cite{25,26} on measuring the
normal Casimir force between sinusoidally corrugated surfaces also allow to
strengthen the constraints \cite{9} by up to an order of magnitude.
Furthermore, it is shown that with increased thickness of Au and Si sectors
of the structured disc in the isoelectronic experiment \cite{21} the obtained
constraints on the Yukawa-type interaction can be strengthened by up to
a factor of 3 over the interaction range from 500\,nm to $1.2\,\mu$m.

The proposed modifications in measurements of the lateral Casimir force \cite{23,24},
normal Casimir force \cite{25,26}, and in the isoelectronic experiment \cite{21}
are also used to derive the prospective constraints on the coupling constants of
axion-like particles to nucleons. According to the results obtained, the modified
experiments will give the possibility to obtain up to a factors of 2.4, 1.7, and 1.7
stronger  constraints, respectively, than  those found in \cite{20} and \cite{22}
from the original measurement data.

The paper is organized as follows. In Sec.~2 the strongest constraints on the
Yukawa-type corrections to Newtonian gravity within the micrometer and
submicrometer interaction range (both already obtained and prospective) are
discussed. Section~3 presents similar results for the axion-to-nucleon coupling
constant in the range of axion masses from 0.1\,meV to 20\,eV. In Sec.~4 the reader
will find the conclusions and discussion.

Throughout the paper units with $\hbar=c=1$ are used.

\section{Constraints on the Yukawa-type corrections to Newtonian gravity}

The Yukawa-type correction to Newton's gravitational potential between
two point masses $m_1$ and $m_2$ placed at the points
$\mbox{\boldmath$r$}_1$ and  $\mbox{\boldmath$r$}_2$ is usually
parametrized as  \cite{3}
\begin{equation}
V^{\rm Yu}(r_{12})=
-\alpha\frac{Gm_1m_2}{r_{12}}\,
e^{-r_{12}/\lambda}
,
\label{eq1}
\end{equation}
\noindent
where $\alpha$ is the dimensionless interaction constant, $G$ is the Newtonian
gravitational constant, and
$r_{12}=|\mbox{\boldmath$r$}_{12}|=|\mbox{\boldmath$r$}_1-\mbox{\boldmath$r$}_2|$.
If the Yukawa interaction arises due to exchange of a scalar particle with mass
$M$ between  two atoms with masses $m_1$ and $m_2$, the quantity $\lambda=1/M$
has a meaning of the Compton wavelength of this scalar particle. If the correction
(\ref{eq1}) arises from multi-dimensional physics, then $\lambda$ is the
characteristic size of a compact manifold generated by the extra dimensions.

It has been known that in the nanometer interaction range the strongest constraints
on $\alpha$ given by the Casimir physics follow \cite{8} from measurements of
the lateral Casimir force between two aligned sinusoidally corrugated Au-coated
surfaces of a sphere and a plate \cite{23,24}. The sphere was made of polystyrene
of density $\rho_s=1.06\,\mbox{g/cm}^3$ and uniformly coated with a layer of Cr
of density $\rho_{\rm Cr}=7.14\,\mbox{g/cm}^3$ and thickness $\Delta_{\rm Cr}=10\,$nm
and then with a layer of Au
of density $\rho_{\rm Au}=19.28\,\mbox{g/cm}^3$ and thickness
$\Delta_{\rm Au}^{\!(s)}=50\,$nm.
The external radius of the sphere was $R=97.0\,\mu$m.
The longitudinal sinusoidal corrugations covering the region of the sphere nearest
to the plate have had an amplitude $A_2=13.7\,$nm and a period $\Lambda=574.7\,$nm.
The corrugated plate was made of hard epoxy with density
$\rho_p=1.08\,\mbox{g/cm}^3$
and coated with a layer of Au of thickness
$\Delta_{\rm Au}^{\!(p)}=300\,$nm.
The sinusoidal corrugations on the plate have had $A_1=85.4\,$nm and the same period
as on a sphere (the latter is a condition for obtaining a nonzero lateral Casimir
force).

The lateral Yukawa force in the experimental configuration described above was found
in \cite{8} by the pairwise summation of potentials (\ref{eq1}) over the volumes
of interacting bodies with subsequent negative differentiation with respect to the
phase shift $\varphi$ between corrugations on a sphere and a plate
\begin{eqnarray}
&&
F_{ps,\rm cor}^{\rm Yu,lat}(a,\varphi)
=8\pi^3G\alpha\lambda^3\Psi_{\rm lat}(\lambda)
\,{e^{-a/\lambda}}
\label{eq2} \\
&&~~~~~~~~~~~~~~~~~~~
\times\frac{A_1A_2}{b\Lambda}\,
{I}_1(b/\lambda)\sin\varphi.
\nonumber
\end{eqnarray}
\noindent
Here, $a$ is the separation distance between the zero levels of corrugations on
a sphere and a plate, $I_n(z)$ is the Bessel function of imaginary argument, and
the following notations are introduced
\begin{eqnarray}
&&
\Psi_{\rm lat}(\lambda)=\left[\rho_{\rm Au}-(\rho_{\rm Au}-\rho_p)
e^{-\Delta_{\rm Au}^{\!(p)}/\lambda}\right]
\label{eq3} \\
&&
\times\left[\rho_{\rm Au}\Phi(R,\lambda)-
(\rho_{\rm Au}-\rho_{\rm Cr})\Phi(R-\Delta_{\rm Au}^{\!(s)},\lambda)
e^{-\Delta_{\rm Au}^{\!(s)}/\lambda}\right.
\nonumber \\
&&
\left.
-(\rho_{\rm Cr}-\rho_{s})
\Phi(R-\Delta_{{\rm Au}}^{(s)}-\Delta_{\rm Cr},\lambda)
e^{-(\Delta_{{\rm Au}}^{(s)}+\Delta_{\rm Cr})/\lambda}\right],
\nonumber\\
&&
\Phi(x,\lambda)= x-\lambda+(x+\lambda)e^{-2x/\lambda},
\nonumber\\
&&
b\equiv b(\varphi)=(A_1^2+A_2^2-2A_1A_2\cos\varphi)^{1/2}.
\nonumber
\end{eqnarray}

In the experiment \cite{23,24}, the lateral Casimir force was independently
measured as a function of the phase shift $\varphi$ between corrugations over
the range of separations $a$ from 120 to 190\,nm. At each separation $a_i$ the
measured maximum amplitude of the lateral Casimir force was achieved at some
phase shift $\varphi_i$ and found in agreement with theoretical predictions
of the exact theory within the limits of the experimental errors
$\Delta_iF_{\rm lat}$. The latter were obtained at the 95\% confidence level.
This means that the Yukawa force (\ref{eq2}) satisfies an inequality
\begin{equation}
\left|F_{ps,\rm cor}^{\rm Yu,lat}(a_i,\varphi_i)\right|
\leq\Delta_iF_{\rm lat}.
\label{eq4}
\end{equation}

\begin{figure}[b]
\vspace*{-4.cm}
\resizebox{0.6\textwidth}{!}{%
\hspace*{-1.7cm} \includegraphics{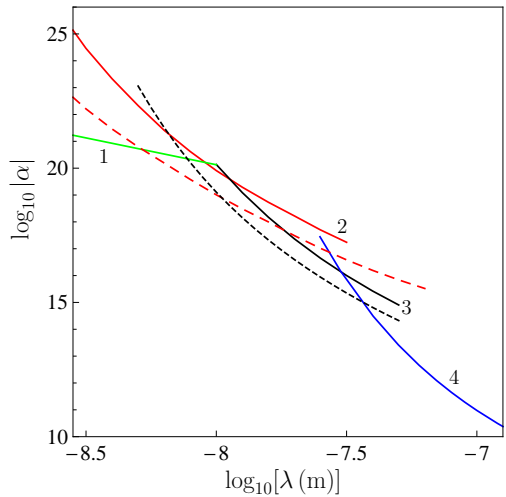}
}
\vspace*{-5.6cm}
\caption{Constraints on the parameters of Yukawa-type correction
to Newton's gravitational law obtained in \cite{10} from the
experiment on neutron scattering (line 1), in \cite{8} from measuring
the lateral Casimir force \cite{23,24} (line 2), in \cite{9} from
measuring the normal Casimir force \cite{25,26} (line 3), and in \cite{21}
from the isoelectronic Casimir-less experiment (line 4). The long-dashed and
short-dashed lines show the prospective constraints obtained in this work
(see the text for further discussion).
The regions of the plane below each line are allowed
and above each line are excluded.}
\label{fg1}       
\end{figure}
In Fig.~\ref{fg1} the constraints on $\alpha$ and $\lambda$ following from
(\ref{eq4}) are reproduced from Fig.~1 of \cite{8} by the solid line 2. Here
and below the regions of the $(\lambda,|\alpha|)$-plane above a line are excluded
and below a line are allowed by the results of respective experiment.
For comparison purposes, the line 1 in Fig.~(\ref{fg1}) shows the strongest
constraints on $\alpha$ and $\lambda$ following from the experiment on neutron
scattering \cite{10}. It can be seen that the constraints of line 2 become
stronger only for $\lambda>9\,$nm. Note that slightly stronger constraints than
those shown by the line 2 were obtained \cite{27} from measurements of the Casimir
force between two crossed cylinders \cite{28}. The experiment \cite{28},
however, suffers from several uncertainties (see \cite{29,30} for details),
which make the deduced results not enough reliable.

Now let us show that the experiment \cite{23,24} on measuring the lateral Casimir force
has a good chance for obtaining stronger constraints at the expense of only minor
modification of the parameters of a setup.
For this purpose, the same corrugation amplitude on a plate is preserved, but the
amplitude on a sphere is increased up to $A_2=25\,$nm. The respective increase in the thickness of an Au coating on a sphere up to $\Delta_{\rm Au}^{\!(s)}=70\,$nm
is made. The corrugation period on a sphere and a plate is decreased
down to $\Lambda=300\,$nm. The sphere radius
$R=100\,\mu$m is chosen almost the same as in the already performed experiment.
Computations of the prospective constraints were done by using
Eqs.~(\ref{eq2})--(\ref{eq4}).
Within the interaction range $\lambda<7\,$nm the strongest constraints on $\alpha$
follow at $a_1=120\,$nm. At this separation the value of the experimental error
$\Delta_1F_{\rm lat}=10\,$pN has been used in agreement with already performed
experiment \cite{23,24}. In a similar way, within the interaction ranges
$7\,\mbox{nm}<\lambda<18\,$nm and $\lambda>18\,$nm
the strongest constraints on $\alpha$ are obtainable at $a_2=125\,$nm and
$a_3=135\,$nm, where $\Delta_2F_{\rm lat}=4.5\,$pN and
$\Delta_3F_{\rm lat}=2.5\,$pN, respectively \cite{23,24}.

The resulting prospective constraints are shown by the long-dashed line
in Fig.~\ref{fg1}. It is seen that they become stronger than the constraints of
line 1, following from the experiments on neutron scattering, for $\lambda>5\,$nm.

Another experiment, used for constraining the Yuka\-wa-type interaction, measured
the normal Casimir force between sinusoidally corrugated surfaces of a sphere and
a plate \cite{25,26}. In this experiment, the polystyrene sphere was coated with
a layer of Cr of thickness $\Delta_{\rm Cr}=10\,$nm, then with a layer of Al
of thickness $\Delta_{\rm Al}=20\,$nm, and finally with a layer of Au
of thickness $\Delta_{\rm Au}^{\!(s)}=110\,$nm. The sphere radius was
$R=99.6\,\mu$m. The parameters of sinusoidal corrugations on the sphere were
$A_2=14.6\,$nm and $\Lambda=570.5\,$nm. The corrugated plate was made of hard
epoxy and coated with a layer of Au of thickness $\Delta_{\rm Au}^{\!(p)}=300\,$nm.
The corrugations on the plate have had the same period, as on a sphere, and
an amplitude $A_1=40.2\,$nm \cite{25,26}. An opposed to experiment on measuring
the lateral Casimir force \cite{23,24}, in this experiment the axes of
corrugations on a sphere and a plate should not be necessary parallel.

For the sake of simplicity, however, here the case of parallel axes of corrugations
is considered. In this case the normal Yukawa force in the experimental configuration
is given by \cite{9}
\begin{eqnarray}
&&
F_{ps,\rm cor}^{\rm Yu,nor}(a)
=-4\pi^2G\alpha\lambda^3\Psi_{\rm nor}(\lambda)
\,{e^{-a/\lambda}}
\label{eq5} \\
&&~~~~~~~~~~~~~~~~~~~~~~~~~~~~~~
\times
{I}_0\left(\frac{A_1-A_2}{\lambda}\right),
\nonumber
\end{eqnarray}
\noindent
where
\begin{eqnarray}
&&
\Psi_{\rm nor}(\lambda)=[\rho_{\,\rm Au}-
(\rho_{\,\rm Au}-\rho_{p})
e^{-\Delta_{\rm Au}^{\!(p)}/\lambda}]
\label{eq6} \\
&&~~
\times\left[
R\rho_{\,\rm Au}-
(\rho_{\,\rm Au}-\rho_{\,\rm Al})(R-\Delta_{\rm Au}^{\!(s)})
e^{-\Delta_{\rm Au}^{\!(s)}/\lambda}\right.
\nonumber \\
&&~~
-(\rho_{\,\rm Al}-\rho_{\,\rm Cr})
(R-\Delta_{\rm Au}^{\!(s)}-\Delta_{\rm Al})
e^{-(\Delta_{\rm Au}^{\!(s)}+\Delta_{\rm Al})/\lambda}
\nonumber \\
&&~~
-(\rho_{\,\rm Cr}-\rho_{s})
(R-\Delta_{\rm Au}^{\!(s)}-\Delta_{\rm Al}-\Delta_{\rm Cr})
\nonumber \\
&&~~~~~~~~~~~~~~~~~~~~~~
\left.\times
e^{-(\Delta_{\rm Au}^{\!(s)}+\Delta_{\rm Al}+\Delta_{\rm Cr})/\lambda}
\right].
\nonumber
\end{eqnarray}

The theoretical predictions for the normal Casimir force calculated using the
scattering theory have been confirmed experimentally within the experimental errors
$\Delta_iF_{\rm nor}$. This means that any additional normal force should satisfy
the inequality
\begin{equation}
\left|F_{ps,\rm cor}^{\rm Yu,nor}(a_i)\right|
\leq\Delta_iF_{\rm nor}.
\label{eq7}
\end{equation}

The constraints on the parameters of Yukawa interaction obtained from
Eqs.~(\ref{eq5})--(\ref{eq7}) are shown by the line 3 in Fig.~\ref{fg1}
reproduced from Fig.~4 of \cite{9}. As is seen in Fig.~\ref{fg1},
the constraints of line 3 become stronger than the constraints of line 2
for $\lambda>11\,$nm. Thus, the line 2 presents the strongest constraints
within only a very narrow interaction interval of 2\,nm width.
At the same time, the proposed here improved experiment on measuring
the lateral Casimir force (the long-dashed line in Fig.~\ref{fg1}) leads to
the strongest constraints up to $\lambda=18\,$nm (an intersection between
the long-dashed line and line 3). As a result, the constraints of the
long-dashed line are stronger than those of the line 2 by up to a factor
of 10 within the interaction range from 5 to 18\,nm.
The maximum strengthening holds at $\lambda=9\,$nm at the intersection of
lines 1 and 2.

By modifying parameters of corrugations, it is possible to strengthen the
constraints of line 3 from measurements of the normal Casimir force between
corrugated surfaces. Here, an increase of the corrugation amplitudes on a
sphere and  a plate up to $A_2=25\,$nm and $A_1=85\,$nm, respectively, is
proposed. The corrugation period is decreased to $\Lambda=300\,$nm.
All the other parameters are remained as presented above in accordance
with \cite{25,26}.

The prospective constraints on the Yukawa-type corrections to Newtonian
gravity, which could be obtained from measurements of the normal Casimir
force between corrugated surfaces with increased amplitudes and decreased
periods of corrugations, can be found by substitution of (\ref{eq5}) and
(\ref{eq6}) in (\ref{eq7}). The strongest constraints were obtained at
$a_1=127\,$nm where the experimental error was
$\Delta_1F_{\rm nor}=0.94\,$pN \cite{9,25,26}.
The derived constraints are shown by the short-dashed line in Fig.~\ref{fg1}.
They become stronger than the constraints of line 1 for $\lambda>7.5\,$nm.

The recently performed isoelectronic experiment is an important breakthrough
in the field. It allowed significant strengthening of the constraints on
Yukawa interaction in the micrometer and submicrometer interaction ranges
\cite{21}. As discussed in Sec.~1, in the isoelectronic experiment the
contribution of the Casimir force is nullified. This is achieved by making
the difference force measurement between a smooth Au-coated sphere and either
Au or Si sectors of the structured disc. A sapphire sphere of density
$\rho_{\rm sap}=4.1\,\mbox{g/cm}^3$
was coated with a layer of Cr of thickness $\Delta_{\rm Cr}=10\,$nm and then
with a layer of Au of thickness $\Delta_{\rm Au}^{\!(s)}=250\,$nm.
The resulting sphere radius was $R=149.3\,\mu$m. The structured disc consisted
of Au and Si sectors ($\rho_{\rm Si}=2.33\,\mbox{g/cm}^3$) of thickness
$D=2.1\,\mu$m and was coated with overlayer of Cr of thickness
$\Delta_{\rm Cr}=10\,$nm and Au of $\Delta_{\rm Au}^{\!(p)}=150\,$nm thickness.
These overlayers nullify the difference of Casimir forces between
a Au-coated sphere
and Au and Si sectors of the disc. They do not contribute to the difference
of Yukawa forces between  a sphere and the sectors.

The difference of Yukawa-type forces in the experimental configuration of
\cite{21} is given by \cite{21,31,32}
\begin{eqnarray}
&&
F_{ps,\rm diff}^{\rm Yu,nor}(a)=-4\pi^2G\alpha\lambda^3Re^{-a/\lambda}
(\rho_{\rm Au}-\rho_{\rm Si})
\label{eq8} \\
&&~~
\times(1-e^{-D/\lambda})
\left[
\rho_{\rm Au}+(\rho_{\rm Cr}-\rho_{\rm Au})e^{-\Delta_{\rm Au}^{\!(s)}/\lambda}
\right.
\nonumber \\
&&~~~~~~
\left.
+(\rho_{\rm sap}-\rho_{\rm Cr})e^{-(\Delta_{\rm Au}^{\!(s)}+\Delta_{\rm Cr})/\lambda}
\right].
\nonumber
\end{eqnarray}
\noindent
Here,  $a$ is the separation distance between the sphere and Au and Ni sectors of the
disc. The experimentally measured separation between the two test bodies is given by
\begin{equation}
z=a-\Delta_{\rm Au}^{\!(p)}-\Delta_{\rm Cr}.
\label{eq9}
\end{equation}

In the experiment \cite{21} no differential force was observed within the minimum
detectable force $F_{\min}(a)$. This means that the difference of Yukawa-type forces
(\ref{eq8}) satisfies the inequality
\begin{equation}
|F_{ps,\rm diff}^{\rm Yu,nor}(a)|\leq
F_{\min}(a).
\label{eq10}
\end{equation}

\begin{figure}[t]
\vspace*{-4.cm}
\resizebox{0.6\textwidth}{!}{%
\hspace*{-1.7cm} \includegraphics{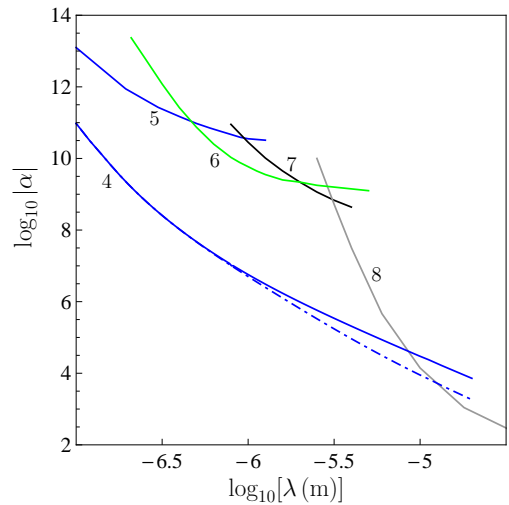}
}
\vspace*{-5.6cm}
\caption{Constraints on the parameters of Yukawa-type corrections
to Newton's gravitational law obtained in the recent \cite{21} and
previous \cite{36} isoelectronic experiments (lines 4 and 5, respectively),
from measuring the difference of lateral forces \cite{37} (line 6),
from the torsion pendulum experiment \cite{38} (line 7), and from the
Cavendish-type experiments \cite{39,40} (line 8).  The dashed-dotted
line shows the prospective constraints obtained in this work
(see the text for further discussion).
The regions of the plane below each line are allowed
and above each line are excluded.}
\label{fg2}       
\end{figure}

The numerical analysis of (\ref{eq8}) and (\ref{eq10}) results \cite{21} in the
line 4 in Fig.~\ref{fg1} which is presented under the same number in Fig.~\ref{fg2}
over a wider interaction range extended to larger values of $\lambda$.
It is seen that the constraints of line 4 become stronger
than the constraints of line 3, derived from measuring the normal Casimir force
between corrugated test bodies, for $\lambda>31\,$nm. Approximately the same left
border of the region, where the constraints of line 4 are the strongest ones, is
given by experiments on measuring the effective Casimir pressure between two
parallel plates by means of
micromachined oscillator \cite{33,34}. The latter constraints are not shown in
Fig.~\ref{fg1} because they are slightly weaker than those shown by the line 3.
Similarly, the constraints following from measurements of the Casimir force between
the smooth surfaces of a sphere and a plate by means of an atomic force
microscope \cite{35} are even weaker \cite{30} and
do not determine the left border of the region where  line 4 indicates the
strongest constraints obtained so far.

{}From Fig.~\ref{fg1} one can also see that
the proposed modification of an experiment on measuring the normal Casimir force
between corrugated surfaces of a sphere and a plate
would lead to the strongest constraints up to
$\lambda=36\,$nm where the short-dashed line intersects the line 4.
By and large the constraints of the short-dashed line are up to an order of
magnitude stronger than the constraints of line 3. The maximum strengthening holds
at $\lambda=11\,$nm.

For comparison purposes, in Fig.~\ref{fg2} the  constraints
on  Yukawa-type corrections to Newtonian gravity following  from
the previous Casimir-less experiment \cite{36} (line 5),
from recent experiment on measuring the difference of lateral forces
\cite{37} (line 6), from measuring the Casimir force using the torsion
pendulum  \cite{38} (line 7), and from the
Cavendish-type experiments \cite{39,40} (line 8) are also shown.
As is seen in Fig.~\ref{fg2}, the isoelectronic experiment (line 4)
provides the most strong constraints over a wide interaction range
$\lambda\leq 8\,\mu$m, and narrows the region of $\lambda$ where the
gravitational constraints have been considered as the strongest ones.
It is seen also that the isoelectronic experiment alone provides
up to a factor of 1000 stronger constraints than several other
experiments using different laboratory setups.

Here, the possibility to further strengthen the constraints on non-Newtonian
gravity following from the isoelectronic experiment \cite{21} is proposed.
For this purpose, it is suggested to increase the thickness of Au and Si sectors
up to $D=10\,\mu$m. The resulting strengthening of the obtained constraints is
determined by the single factor in Eq.~(\ref{eq8}) containing the quantity
$\exp(-D/\lambda)$. These constraints are shown by the dashed-dotted line in
Fig.~\ref{fg2}. As a result, the prospective isoelectronic experiment would
present the strongest constraints in the wider interaction
range $31\,\mbox{nm}<\lambda<12\,\mu$m. The largest strengthening up to a factor
of 3 will be achieved at $\lambda=8\,\mu$m.

\section{Constraints on the coupling constants of axion-like particles to nucleons}

The experiments of the Casimir physics discussed above allow constraining the
coupling constants of axion-like particles to a proton and
a neutron if an interaction
via the pseudoscalar Lagrangian is assumed. In this case the spin-independent
effective potential between two nucleons at the points
$\mbox{\boldmath$r$}_1$ and  $\mbox{\boldmath$r$}_2$ is caused by the exchange of
two axions \cite{14,41,42}
\begin{equation}
V_{kl}^{a}(r_{12})=-
\frac{g_{ak}^2g_{al}^2}{32\pi^3m^2}\,
\frac{m_a}{r_{12}^2}\,
K_1(2m_ar_{12}).
\label{eq11}
\end{equation}
\noindent
In this equation,  the coupling constants between axion and proton ($k,\,l=p$)
and neutron ($k,\,l=n$) are notated $g_{ak}$ and $g_{al}$,
$m$ and $m_a$ are the mean nucleon and axion masses, respectively,
and $K_1(z)$ is the modified Bessel function of the second kind.

Let us begin with an experiment on measuring the lateral Casimir force between
corrugated surfaces on a sphere and a plate \cite{23,24} briefly discussed in
Sec.~2. The additional lateral force due to two-axion exchange in the
experimental configuration was found in \cite{20}. At the phase shift
$\varphi=\pi/2$ between corrugations the amplitude of this force is
equal to \cite{20}
\begin{eqnarray}
&&
\max|F_{ps,\rm cor}^{a,\rm lat}(a)|=
\frac{\pi^2 RC_{\rm Au}}{m_am^2m_{\rm H}^2}\,
\frac{A_1A_2}{\Lambda\sqrt{A_1^2+A_2^2}}
\label{eq12} \\[1mm]
&&~~
\times
\int_{1}^{\infty}\!\!\!du\frac{\sqrt{u^2-1}}{u^3}
e^{-2m_aua} I_1\left(2m_au\sqrt{A_1^2+A_2^2}\right)
\nonumber \\[1mm]
&&~~~~~
\times
(1-e^{-2m_au\Delta_{\rm Au}^{\!(p)}})
\left[
\vphantom{e^{-2m_au\Delta_{\rm Au}^{\!(1)}}}
C_{\rm Au}+(C_{\rm Cr}-C_{\rm Au})
\right.
\nonumber \\[1mm]
&&~~~~\left.
\times e^{-2m_au\Delta_{\rm Au}^{\!(s)}}
-C_{\rm Cr}
e^{-2m_au(\Delta_{\rm Au}^{\!(s)}+\Delta_{\rm Cr})}
\right].
\nonumber
\end{eqnarray}
\noindent
Here, $m_{\rm H}$ is the mass of atomic hydrogen and the coefficient
$C$ for any material $k$ ($k=\rm Au,\,Cr$ etc.) is defined as
\begin{equation}
C_k=\rho_k\left(\frac{g_{ap}^2}{4\pi}\,\frac{Z_k}{\mu_k}+
\frac{g_{an}^2}{4\pi}\,\frac{N_k}{\mu_k}\right),
\label{eq13}
\end{equation}
\noindent
where $Z_k$ and $N_k$ are the number of protons and the mean number of neutrons
in an atom or a molecule of the material under consideration,
$\mu_k=m_k/m_{\rm H}$, and $m_k$ is the mean atomic (molecular) mass.
Specifically,
$Z_{\rm Au}/\mu_{\rm Au}=0.40422$, $Z_{\rm Cr}/\mu_{\rm Cr}=0.46518$,
$N_{\rm Au}/\mu_{\rm Au}=0.60378$, and $N_{\rm Cr}/\mu_{\rm Cr}=0.54379$
\cite{3}. Note that poly\-sty\-re\-ne and hard epoxy contribute negligibly small
to the force due to two-axion exchange, and these contributions are omitted
in (\ref{eq12}).

\begin{figure}[t]
\vspace*{-4.cm}
\resizebox{0.6\textwidth}{!}{%
\hspace*{-1.7cm} \includegraphics{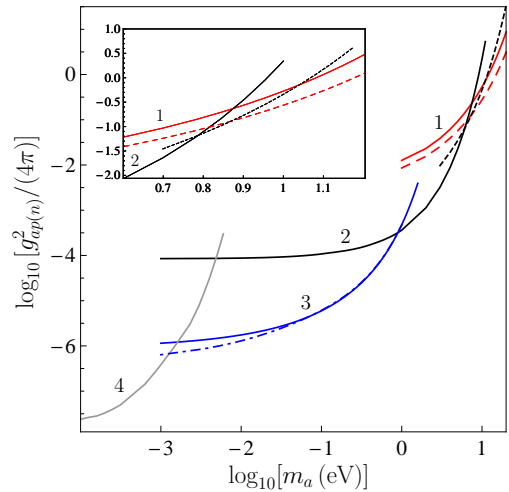}
}
\vspace*{-5.6cm}
\caption{Constraints on the coupling constants of axion-like
particles to nucleons  obtained in \cite{20} from
measuring the lateral Casimir force \cite{23,24} (line 1),
in \cite{19} from measuring the effective Casimir pressure
\cite{33,34} (line 2), in \cite{22} from the
isoelectronic experiment \cite{21} (line 3), and in \cite{16}
from the Cavendish-type experiment \cite{15} (line 4).
The long-dashed, short-dashed and dashed-dotted lines show
the prospective constraints obtained in this work
(see the text for further discussion).
In the inset, the range of larger axion masses is presented 
on an enlarged scale.
The regions of the plane below each line are allowed
and above each line are excluded.
}
\label{fg3}       
\end{figure}

The constraints on the $g_{an}$ and $g_{ap}$ have been found by substitution of the
force amplitude (\ref{eq12}) in (\ref{eq4}) in place of
$F_{ps,\rm cor}^{\rm Yu,lat}$ (see Fig.~2 of \cite{20}). They are shown by the line~1
in Fig.~\ref{fg3} under a condition  $g_{an}=g_{ap}$. As in the case of Yukawa
interaction, the regions above each line in Fig.~\ref{fg3} are excluded and below
each line are allowed.

Now let us consider the modified setup with increased
corrugation amplitudes and decreased period as indicated in Sec.~2. By making
computations with the help of (\ref{eq12}), (\ref{eq13}) and (\ref{eq4}),
where $F_{ps,\rm cor}^{\rm Yu,lat}(a)$ is replaced with
$\max|F_{ps,\rm cor}^{a,\rm lat}(a)|$, one
arrives to stronger constraints on $g_{an}=g_{ap}$ shown by the long-dashed line
in Fig.~\ref{fg3}. They are shown also in an inset to Fig.~\ref{fg3} on an
enlarged scale.

In one more experiment discussed in Sec.~2 the normal Casimir force between
the sinusoidally corrugated surfaces has been measured \cite{25,26}.
The additional normal force arising in this experimental configuration due
to two-axion exchange was found in \cite{20}. It is given by
\begin{eqnarray}
&&
F_{ps,\rm cor}^{a,\rm nor}(a)=-\frac{\pi RC_{\rm Au}}{2m_am^2m_{\rm H}^2}
\int_{1}^{\infty}\!\!\!du\frac{\sqrt{u^2-1}}{u^3}
\label{eq14} \\[1mm]
&&~~~
\times  e^{-2m_aua}
I_0\left(2m_au(A_1-A_2)\right)(1-e^{-2m_au\Delta_{\rm Au}^{\!(p)}})
\nonumber \\[1mm]
&&~~~~~
\times\left[C_{\rm Au}+(C_{\rm Al}-C_{\rm Au})
e^{-2m_au\Delta_{\rm Au}^{\!(s)}}\right.
\nonumber \\[1mm]
&&~~~~~~~
+(C_{\rm Cr}-C_{\rm Al})
e^{-2m_au(\Delta_{\rm Au}^{\!(s)}+\Delta_{\rm Al})}
\nonumber \\[1mm]
&&~~~~~~~
\left.
-C_{\rm Cr}
e^{-2m_au(\Delta_{\rm Au}^{\!(s)}+\Delta_{\rm Al}
+\Delta_{\rm Cr})}\right].
\nonumber
\end{eqnarray}
\noindent
For Al one has \cite{3}
$Z_{\rm Al}/\mu_{\rm Al}=0.48558$ and $N_{\rm Al}/\mu_{\rm Al}=0.52304$.

The constraints on $g_{ap(n)}$ have been found by substituting (\ref{eq14}) in
(\ref{eq7}), where the force $F_{ps,\rm cor}^{\rm Yu,nor}$ was replaced with
$F_{ps,\rm cor}^{a,\rm nor}$ (see Fig.~1 of \cite{20}). They turned out to be
weaker than the combined constraints of line 1 in Fig.~\ref{fg3}, following
from measurements of the lateral Casimir force, and of line 2 obtained \cite{19} from measurements of the effective Casimir pressure by means of micromachined
oscillator \cite{33,34}. Because of this, the constraints of \cite{20} are not
reproduced in Fig.~\ref{fg3}. However, with increased corrugation amplitudes and
decreased period, as proposed in Sec.~2, the stronger constraints can be
obtained. They are found from (\ref{eq7}) with the above replacement
and (\ref{eq14}). The derived constraints are shown by the
short-dashed line in Fig.~\ref{fg3}, and on an enlarged scale in the inset
to this figure.

As is seen in Fig.~\ref{fg3}, the proposed experiment on measuring the normal
Casimir force between corrugated surfaces (the short-dashed line) allows to
strengthen the constraints on $g_{ap(n)}$ in the region of axion masses from
6\,eV to 11\,eV. The largest strengthening by a factor of 1.7 holds for
$m_a=8\,$eV. The constraints of the long-dashed line, following from the
proposed experiment on measuring the lateral Casimir force, are stronger than
those of line 2 and of the short-dashed line  for $m_a>6.5$ and 7.5\,eV,
respectively. In so doing, the strengthening by the factors of 2 and 2.4,
as compared to the line 1, are reached for $m_a=10$ and 15\,eV, respectively.
(At the moment the line 1 indicates the strongest constraints for
$m_a>8\,$eV.)

The line 3 in Fig.~\ref{fg3} shows the constraints on  $g_{ap(n)}$  obtained
\cite{22} from the recent isoelectronic Casimir-less experiment \cite{21}.
In the configuration of this experiment the difference of additional forces
due to two-axion exchange is given by \cite{22}
\begin{eqnarray}
&&
|\Delta F_{ps,\rm diff}^{a,\rm nor}(a)|=
\frac{\pi }{2m_am^2m_H^2}(C_{\rm Au}-C_{\rm Si})
\label{eq15}\\
&&~~
\times\!\!
\int_{1}^{\infty}\!\!du\frac{\sqrt{u^2-1}}{u^3}e^{-2m_aua}\left(1-e^{-2m_auD}\right)
X(m_au),
\nonumber
\end{eqnarray}
\noindent
where
\begin{eqnarray}
&&
X(z)= C_{\rm Au}
\left[\chi(R,z)-e^{-2z\Delta_{\rm Au}^{\!(s)}}
\chi(R-\Delta_{\rm Au}^{\!(s)},z)\right]
\label{eq11} \\
&&
+C_{\rm Cr}e^{-2z\Delta_{\rm Au}^{\!(s)}}
\left[\vphantom{e^{-2z\Delta_{\rm Cr}}}
\chi(R-\Delta_{\rm Au},z)\right.
\nonumber \\
&&~~~~~~~~\left.
-e^{-2z\Delta_{\rm Cr}}
\chi(R-\Delta_{\rm Au}^{\!(s)}-\Delta_{\rm Cr},z)\right]
\nonumber \\
&&
+C_{\rm sap}e^{-2z(\Delta_{\rm Au}^{\!(s)}+\Delta_{\rm Cr})}
\chi(R-\Delta_{\rm Au}^{\!(s)}-\Delta_{\rm Cr},z).
\nonumber
\end{eqnarray}
\noindent
and
\begin{equation}
\chi(r,z)=r-\frac{1}{2z}+e^{-2rz}\left(r+
\frac{1}{2z}\right).
\label{eq16}
\end{equation}
\noindent
Note that for Si and sapphire one obtains \cite{3}
$Z_{\rm Si}/\mu_{\rm Si}=0.50238$, $N_{\rm Si}/\mu_{\rm Si}=0.50628$,
and
$Z_{\rm sap}/\mu_{\rm sap}=0.49422$,
 $N_{\rm sap}/\mu_{\rm sap}=0.51412$.

The constraints on the coupling constant of axion-like particles to
nucleons are recalculated here by substituting (\ref{eq15}) in place of
$|\Delta F_{ps,\rm diff}^{\rm Yu,nor}|$ in (\ref{eq10}) with increased
thickness $D$ of Au and Si sectors, as proposed in Sec.~2.
The obtained constraints are shown by the dashed-dotted line in Fig.~\ref{fg3}.
For comparison purposes, the line 4 reproduces constraints on $g_{ap(n)}$
obtained \cite{16} from the gravitational experiment of Cavendish type \cite{15}.
As is seen in Fig.~\ref{fg3}, the constraints of line 3 are the strongest
ones in the range of axion masses from 1.7\,meV to 0.9\,eV. With increased
thickness of Au and Si sectors,  stronger constraints of the dashed-dotted
line could be obtained over the range of $m_a$ from 1.3\,meV to 40\,meV.
The maximum strengthening by a factor of 1.7 holds at $m_a=1.7\,$meV.

\section{Conclusions and discussion}

In the foregoing the strongest constraints on the Yukawa-type corrections
to Newtonian gravity and coupling constants of axion-like particles to
nucleons following from the Casimir physics are collected. Minor modifications
is respective experimental configurations are proposed allowing further
strengthening of the obtained const\-ra\-ints.
Specifically, it is shown that if one preserves the corrugation amplitude on
a plate ($A_1\approx 85\,$nm), but increases it on a sphere from $A_2=13.7\,$nm
to $A_2=25\,$nm, and decreases the period of corrugations $\Lambda$ from 574\,nm
to 300\,nm, the constraints on non-Newtonian gravity, following from measurements
of the lateral Ca\-si\-mir force, become stronger up to a factor of 10.
Similar modifications in the setup for measuring the normal Casimir force
between corrugated surfaces also result in up to an order of magnitude stronger
constraints. An increase of thickness $D$ of Au and Si sectors in the recent
isoelectronic experiment (from 2.1 to $10\,\mu$m), proposed in this paper,
results in up to a factor of 3 stronger constraints on non-Newtonian gravity
over a wide interaction range.

At the moment, the strongest constraints on the Yukawa-type corrections to
Newtonian gravity for \hfill \\
$\lambda<9\,$nm follow from the experiments on neutron
scattering \cite{10}, for $9\,\mbox{nm}<\lambda<11\,$nm from measuring the
lateral Casimir force, for $11\,\mbox{nm}<\lambda<31\,$nm from measuring the
normal Casimir force between corrugated surfaces \cite{25,26},
for $31\,\mbox{nm}<\lambda<8\,\mu$m from the isoelectronic experiment \cite{21},
and for larger $\lambda$ from the Cavendish-type experiment \cite{39,40}.
Thus, it is shown that the constraints of \cite{21} are the strongest ones in
a wider region of $\lambda$ than is indicated in \cite{21}.
If the suggested modifications will be implemented, the strongest constraints
for $\lambda<5\,$nm will follow from  neutron scattering,
for $5\,\mbox{nm}<\lambda<10.5\,$nm from the proposed measurements of the
lateral Casimir force,
for $10.5\,\mbox{nm}<\lambda<36\,$nm from the proposed measurements of the
normal Casimir force,
for $36\,\mbox{nm}<\lambda<12\,\mu$m from the modified isoelectronic experiment,
and for $\lambda>12\,\mu$m from gravitational experiments.

The proposed modifications of the test bodies in measurements of the lateral and
normal Casimir forces, and in the isoelectronic experiment also allow to
streng\-then constraints on the coupling constant of axion-like particles to
nucleons. At the moment, the strongest constraints for
$1\,\mu\mbox{eV}<m_a<1.7\,$meV follow \cite{16} from the Cavendish-type
experiment \cite{15},
for $1.7\,\mbox{meV}<m_a<0.9\,$eV from the isoelectronic experiment \cite{21,22},
for $0.9\,\mbox{eV}<m_a<8\,$eV from measuring the effective Ca\-si\-mir pressure
\cite{19,33,34},
for $m_a>8\,$eV from measuring the lateral Casimir force
\cite{20,23,24}.
If the suggested experiments will be realized, the strongest constraints
for $1.3\,\mbox{meV}<m_a<0.9\,$eV will follow from the isoelectronic experiment,
for $0.9\,\mbox{eV}<m_a<6\,$eV from measuring the effective Casimir pressure,
for $6\,\mbox{eV}<m_a<7.5\,$eV from the proposed measurements of the normal
Casimir force, and
for $m_a>7.5\,$eV from the proposed measurements of the lateral Casimir force.

Thus, the Casimir physics already resulted in strong laboratory constraints
on the non-Newtonian gravity and axion-like particles. As shown above,
the proposed minor modifications of already performed experiments make
possible obtaining even stronger constraints. In future, more radical
improvements in the laboratory setups may be employed (for instance,
by using the test bodies with aligned nuclear spins \cite{43} or
a large area force sensor suggested
for constraining chameleon interactions \cite{44}).

\section*{Acknowledgement}
The author is grateful to R.\ S.\ Decca for
sending the numerical data for line 4 in Figs.~\ref{fg1} and \ref{fg2}
and to V.\ M.\ Mostepanenko for useful discussions.

\end{document}